# Multi-stage optimisation towards transformation pathways for municipal energy systems

## Paul Maximilian Röhrig, Nils Körber, Julius Zocher, Andreas Ulbig

Institute for High Voltage Equipment and Grids, Digitalization and Energy Economics (IAEW), RWTH Aachen University, 52056 Aachen, Germany

Fraunhofer-Zentrum Digitale Energie, 52068 Aachen, Germany

**Contact:**
Paul Maximilian Röhrig, M.Sc. | Tel.: +49 241 80- 94937| m.roehrig@iaew.rwth-aachen.de | www.iaew.rwth-aachen.de

**ABSTRACT**

An essential facet of achieving climate neutrality by 2045 is the decarbonization of municipal energy systems. To accomplish this, it is necessary to establish implementation concepts that detail the timing, location, and specific measures required to achieve decarbonization. This restructuring process involves identifying the measures that offer the most compelling techno-economic and ecological advantages. In particular, measures that contribute to the interconnection of energy vectors and domains, e.g. heating, cooling, and electricity supply, in the sense of decentralized multi-energy systems are a promising future development option. Due to the high complexity resulting from a multitude of decision options as well as a temporal coupling across the transformation path, the use of optimization methods is required, which enable a bottom-up identification of suitable transformation solutions in a high spatial resolution. For the design of reasonable concepts, we develop a multistage optimization problem for the derivation of transformation pathways in the context of a multi-location structure, expansion, and operation problem. The results show that the heat supply in the future will mainly be provided by heat pumps with a share of 60%. It can also be shown that an early dismantling of the gas network will lead to the need for transitional technologies such as pellet heating. Overall, the conversion of the municipal energy system can significantly reduce emissions (97%).

## 1. INTRODUCTION

### *Motivation*

The goal of various states to achieve climate neutrality in the coming years leads to the requirement to reduce emissions in all energy sectors. A large part of these emissions (13%) is caused by the energy supply of buildings, which is composed of the final energy forms electricity, cooling, and heat [1].

Possible approaches at the building level to achieve zero-carbon buildings include reducing final energy consumption, such as renovating the building envelope and expanding local renewable energy generation by installing photovoltaic roof systems. Another measure is the replacement of fossil technologies, such as gas condensing boilers, with alternatives, such as heat pumps [2]. There are both technical and economic interdependencies between the various measures. The building level also interacts with its higher level, the so-called municipal energy system, which provides the infrastructure for electricity, gas, and heating networks. In this context, the grid can be both a driver and a barrier to the energy transition. A rapid dismantling of the gas grid forces end customers to switch from existing fossil gas plants.

For Germany, this would mean having 23 years to meet its climate targets. Due to the longevity of technologies at the building and city levels, targeted decisions must be made on when, where, and what to convert. The necessary implementation pathways should be as robust as possible to deal with future uncertainties (e.g., energy price development). The paths can be defined by intermediate targets, which represent a recommendation for action. These uncertainties coupled with the need to adapt the existing system can lead to path dependencies. For example, the conversion of the gas network to a hydrogen network can lead to stranded investments if they become obsolete in the future due to other technologies [3]. In addition, implementation is limited in time since, on the one hand, each implementation process requires a certain amount of time, and on the other hand, processes cannot run in parallel indefinitely. For example, the renovation rate, i.e. the rate of buildings that can be renovated annually, is currently 1%, which means that in terms of time, not every building can achieve complete renovation by the target year set by the German government [4].

*Literature Review*

An approach to derive transformation pathways in the municipal context is needed. Existing approaches, such as MANGO [5] focus on deriving year-specific developments at the building level and neglect interactions with the municipal energy system. Other works like OSeMOSYS [6] offer approaches to derive supra-regional developments but abstract from the individual decisions per building. This form of aggregation thus neglects local georeferenced contexts of a municipal energy system, which can lead to incorrect decisions.

*Contributions*

This paper introduces a method for deriving a robust transformation path for a municipal energy system. It allows for buildingspecific consideration, with broad technical coverage of possible solutions across different energy sources. The developed method uses a myopic approach to derive technology penetrations per building and the time period under consideration to map interactions with the superposition layer.

*Organization*

The paper is organized as follows: In Section II, an overview of the methodological flow is given, and the core elements of the procedure are presented for this purpose. Based on this, an outlook on an exemplary application of the method is presented in section III.

## 2. METHODOLOGY

*Overview*

The developed model corresponds to multi-stage optimization based on a mixed linear optimization model for the derivation of techno-economic/ecological design options for the heating, cooling, and electricity supply of the municipal energy system, including the individual decision of building level in a bottom-up modeling approach. The method is based on a further development of the approach from [7]. The aim of the existing method is to derive the technology penetration for a future greenfield district energy system. For this purpose, an optimization method is used, which is able to perform an expansion and operation decision with a user-centered perspective with the target variables of economic and ecological minimization. The model's level of detail covers a wide range of spatial, temporal, technical, and content-related aspects (tab. I/II).

TABLE I. AVAILABLE TECHNOLOGIES

| Heat sector | | |
|---|---|---|
| ▪ Solarthermal | ▪ Oil-heating | ▪ Pellet-heating |
| ▪ Gas condensing boiler | ▪ Ground source heat pump | ▪ Woodchip heating |
| ▪ Direct current heating | ▪ Air source heat pump | ▪ Buffer tank |
| ▪ Fuel cell | ▪ Mirco combined heat and power plant | ▪ Heat exchanger |
| **Cooling sector** | | |
| ▪ Air conditioner | ▪ Heat exchanger | |
| **Electricity sector** | | |
| ▪ Fuel cell | ▪ Mirco combined heat and power plant | ▪ Photovoltaics |
| ▪ Lio-Battery | ▪ Grid connection | |

TABLE II. AVAILABLE REFURBISHMENT

| **Refurbishment** | | | |
|---|---|---|---|
| ▪ Roof | ▪ Wall | ▪ Window | ▪ Cellar |

*Extention*

1) Requirements for the existing model: In order to derive realistic development of the transformation path of a municipal energy system, the existing greenfield approach has to be extended to a brownfield approach. To perform the extension, existing technologies are considered in the optimization problem, so that besides an expansion of technologies, the status quo technology can still be used, when its expected lifetime is not reached in the relevant period under consideration. A replacement of technologies before their expected end of lifetime is considered as opportunity costs. This leads to a deterioration of the target value in the case of dismantling in comparison to a green-field approach. The objective function consists of the summed costs for expansion, subvention, operation, deconstruction, and residual value of different technologies (T) per building (B) (eq.1).

$$\min \sum_{b \in B} \sum_{t \in T} (a_{b,t}^{CAPEX} - a_{b,t}^{CAPEX\,SUB} + a_{b,t}^{OPEX} + a_{b,t}^{DECONST} + a_{b,t}^{RESIDUAL\,VALUE}) \tag{1}$$

Further, existing refurbishment statuses and technologies must be integrated into the model, and the following are the most important extensions.

$$\sum_{t \in T_{IINV \in T}} x_t^{dim} \leq A_b^{Roof} - \sum_{t \in T_{SQ \subset T}} x_t^{dim} \quad \forall b \in B \tag{2}$$

The formula 2 limits the maximum expansion of roof technologies to the available area minus the share of already existing systems. Similar limitation is used for the other technologies in terms of power class. In the model, a distinction is made between 15 refurbishments and an unrefurbished variant. In order to represent existing investments, all variants that can no longer be achieved are deactivated. This allows the model to remain modular (eq. 3).

$$x_b^{bin_{ir}} = 0 \quad \forall b \in B, ir \in Inadmissible\ Refurb. Variants \tag{3}$$

In the case of existing technology, the model can decide between continued use of the technology and dismantling (eq. 4). The objective function for continued use is the negative residual value of the technology, while the objective function for dismantling is the resulting costs. If dismantling is chosen, the objective function is worsened by the unused residual value plus the dismantling costs.

$$x_{b,t}^{bin_{sq}} + x_{b,t}^{bin_{dec}} = 0 \quad \forall b \in B, t \in T \tag{4}$$

In addition to deciding whether and which technology should be used, it is important to consider the time component of the implementation measures (5). This means that measures cannot be carried out in parallel indefinitely and each measure requires a certain amount of time for implementation. The implementation rate is largely determined by the available craftsmen. Renovation measures are more complex and time-consuming than the installation of heating technologies ([4], [2] [8],[9], [10], [8]). Examples for the duration of specific implementation measures are:

- Installing a new heating system: a few days
- Renovating a facade: several weeks to months
- Building a new house: several months to years

$$\sum_{t \in T_{IINV \in T}} x_t^{bin} \leq max.\ retrofit\ in\ a\ period\ under\ consideration \quad \forall b \in B \tag{5}$$

2) Model layout: **Inputs**: The derivation of a robust building-level transformation path uses a serial, rolling, linearized optimization as described in figure 1. The input data of the method is a digital energy twin of the existing system, which contains the characteristic demand structure per building and potentials for renewable energies on a central as well as building specific level. In Addition, it describes which technical facilities are available so far and which will be available in the future. Furthermore, the scenario framework of the study is defined, which specifies the time of calculation and the exogenous factors (energy prices, grid charges, resource availability, etc.) for the system in the future.

**Pre-Processing**: The serial determination of technology penetration divides the study framework into sub-periods that can be dynamically defined in the input data. Within each of these time periods, building age and technologies are determined incrementally for the study period (eq.6). Additionally, the solution space of the building is checked, e.g. whether a technology can provide the necessary peak power.

$$time_{b,inst.year_t} + time_{b,lifetime_t} \leq time_{target_{year}} \quad \forall b \in B, t \in T \tag{6}$$

**Optimization**: Within the optimization core, a deterministic decision with full information is made for each building with the degrees of freedom of the expansion of facilities, the flexible operation, the deconstruction of the facility and the further use of existing facilities if available and within their lifetime (fig.1). Due to the high complexity of the existing model [7], decision processes that cause cross-building couplings are removed from the optimization process and decided separately. This includes, among other things, the time constraint of decisions, the so-called time heuristic. This is made retroactively on the basis of the optimization result, which was previously unconstrained. Within the optimization, there are certain conditions that must be met. These include the configuration of the plants to ensure the future energy demand, as well as compliance with technical restrictions. In addition, regulatory requirements must also be taken into account, such as the prohibition of certain technologies. The interconnection between the plants is also taken into account to enable differentiated subsidies and to determine the degree of self-consumption in detail. The result of the optimization provides a statement about a possible refurbishment strategy and the appropriate technology decision for electricity, heating and cooling per building. In addition, the results include costs (CAPEX, OPEX, residual value; deconstruction costs), emissions (commissioning, operation, deconstruction) and the respective operating results (e.g. generation of the PV system is used to x% for self-supply). This unrestricted result must be finally evaluated in the follow-up regarding, resources and overall feasibility.

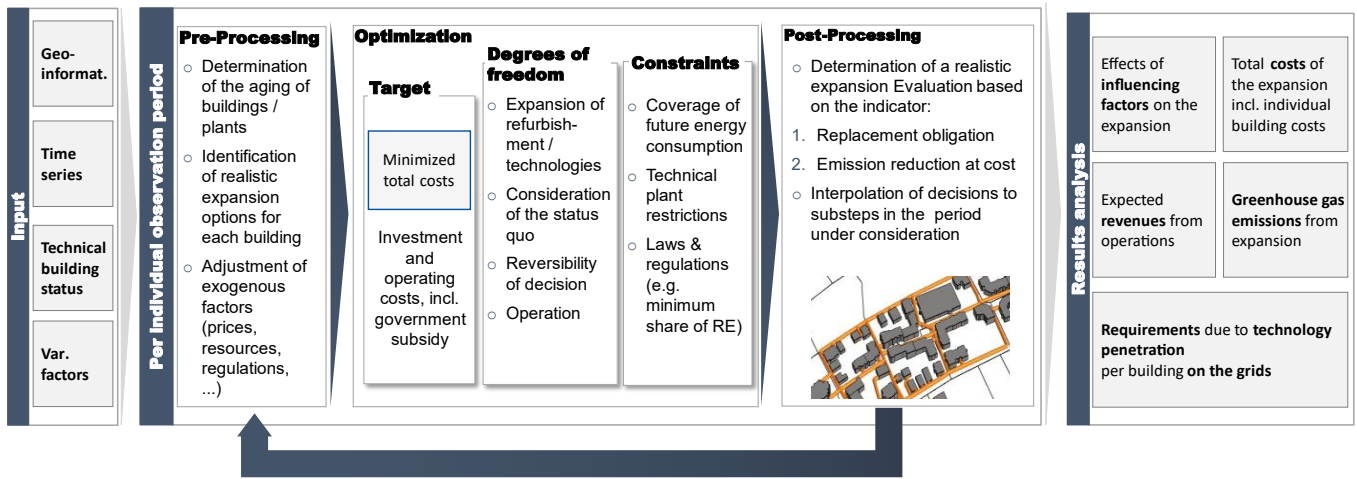

Figure 1.  Schematic representation of the optimization procedure

**Post-Processing**: To determine the feasibility of the outcome, decisions are divided into two categories: First, building retrofits (Refurbishments/Technology Invests) that are unavoidable, such as new construction of buildings and thus the initial installation of a technology; and second, voluntary retrofit based on the reduction potential of costs and emissions compared to existing building system. The assumption is that buildings tend to implement a measure first if it has the greatest impact on one or both of the decision variables ([11], [12],[13], [14]).

The group of voluntary retrofit measures is divided into two subgroups: energy renovations and plant conversions. Energy renovations are limited differently than plant conversions due to their duration and complexity. If a retrofit cannot be performed due to limited resources (e.g., craftsmen), the building can be further optimized between the facility retrofit options and maintaining the status quo technologies. The number of buildings that only want to perform a plant conversion is also limited due to resource scarcity. If a building cannot perform a retrofit, further optimization will still be performed without changing the degree of freedom of retrofits or plant engineering. This is used to evaluate the operating result under the dynamic scenario frameworks for the period under consideration. The described procedure is shown in Figure 3.

**Outputs**: In addition to deriving realistic expansion decisions according to these categories, the determined technology penetration can be used for a grid calculation on the one hand, and on the other hand, the determined technology penetration for the year under study can be used as an input variable for the subsequent study. The output of the model is a transformation path of a municipal energy system for different years, including the total cost of expansion for each building, as well as the expected revenue from the sale of surplus production in operation and emissions or emission savings as a delta between status quo and target year.

3) Simplification techniques: A number of simplification techniques are used to reduce the complexity of the optimization model, which are presented in the following. Within the framework, the investigation period of, for example, 2023-2045 is broken down into smaller time periods, e.g. five-year steps, to reduce the problem complexity while maintaining the representation of dynamic uncertainties.

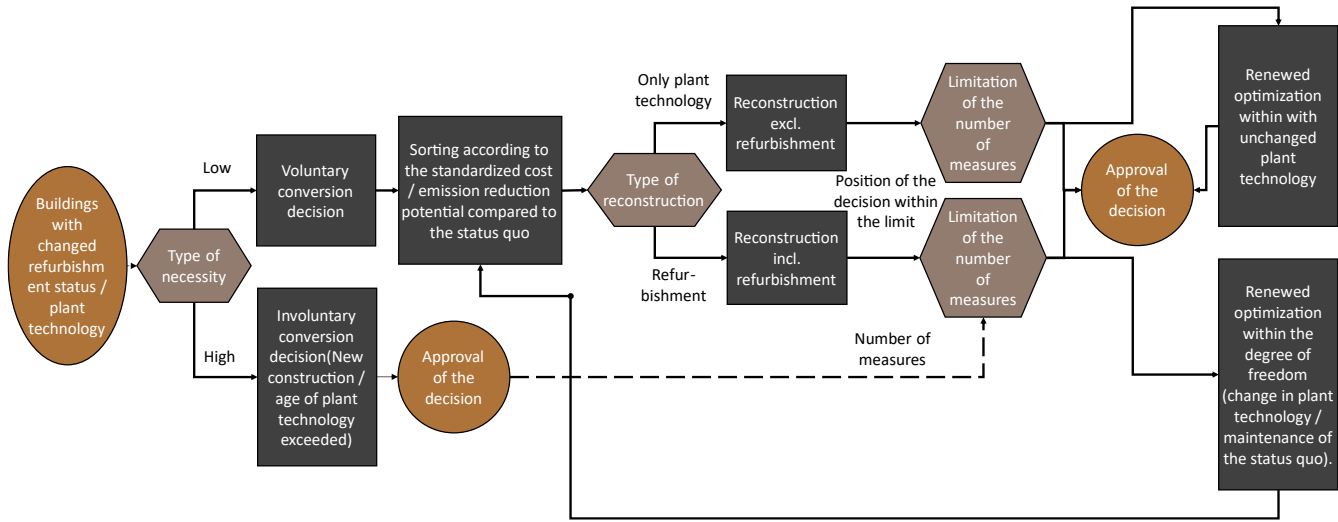

Figure 2.  Schematic representation of the heuristic

A further simplification is assumed for the optimization; only the target year 2030 is optimized in this representation, with 2025 forming the status quo. In the optimization a yearly time range with a resolution of 60min is used to consider the plant operation which could be further simplified by means of time series aggregation, in order to further improve the calculation speed. An ideal target image without external restriction is optimized, using the time heuristic (fig. 2) this is now restricted to realistic solutions in this time range. By means of interpolation and ranking based on least cost, the individual measures can now be assigned years, when they are ideally implemented, thus it is possible to represent the time range in a more deferenced manner. The procedure is based on the description of [5].

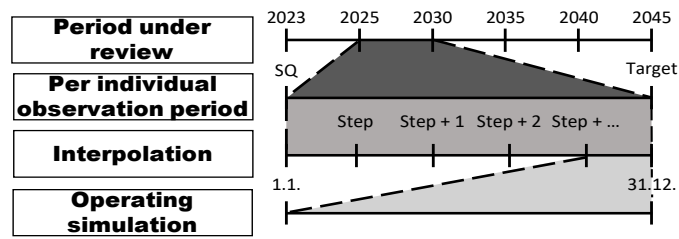

Figure 3.  Schematic representation of the time relations

The resulting change in the technology penetration of the buildings can then be calculated by interpolating to sub-steps between the status quo and the outcome. In this process, an implementation year is assigned to each decision based on the previously mentioned reduction value, which leads to an increase in the level of detail of the transformation path (fig. 2). The results can now be used as an initial state for a rolling optimization process to derive a transformation path for the total period under consideration.

## 3.    RESULTS

In an exemplary case study, the transformation path of a municipal energy system is analyzed. For this purpose, the object of study is presented in section III-A and the results of the modeling are analyzed in section III-B. The technology penetration is calculated for the years 2023, 2025, 2030, 2035, 2040, and 2045, assuming a maximum renovation rate of 2.0% [4].

### *Object of investigation*

This paper examines a fictitious, representaive municipal energy system with urban development. With an area of 4.4 km2 and a population of about 35.000 inhabitants, it is a representative example of a typical, heterogeneous municipality. In total, the area comprises 3127 buildings which are distributed over different building types and uses. Residential buildings are the dominant group with 88.5%, followed by commercial buildings and public facilities.The average age of buildings in the region is between 1933 and 1950 years, resulting in an average heating age of 10 years. The energy demand for the region is 120.8 GWh for electrical energy and 587.5 GWh for thermal energy. This demand is met by various energy sources, including electricity, gas, oil, heatnetwork and renewable energy (fig. 4). Heat generation is dominated by gas condensing boilers with 84%. The following self-defined scenario describes an exemplary price development for the future and is used for the calculations in this paper, see Table III. The scenario describes a decrease in electricity purchase costs as well as an increase of the costs of fossil and alternative energy sources. In optimizing the transformation path, the model examines various measures for increasing energy efficiency and reducing greenhouse gas emissions, including the use of energy-efficient renovations, the expansion of renewable energy and the replacement of fossil-emitting technology. In the area is already an existing heating network which can supply only isolated streets. It is assumed in the optimization that the network has fully tapped its heat potential sources and therefore cannot be expanded further.

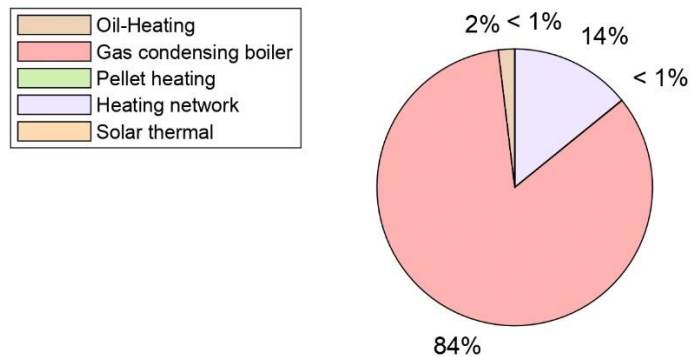

Figure 4. Distribution of technologies in heating demand at 2023

TABLE II. SCENARIO ASSUMPTIONS

| Years [a] | Price [€ct/kWh] | | | | |
|---|---|---|---|---|---|
|  | Electricity | Gas | Oil | Pellets | $CO_2$ certificate |
| 2023 | 49.39 | 18.64 | 6.81 | 5.51 | 80 |
| 2025 | 40.66 | 13.94 | 9.35 | 7.70 | 90 |
| 2030 | 31.62 | 14.63 | 12.61 | 10.97 | 130 |
| 2035 | 25.40 | 15.41 | 15.42 | 14.22 | 150 |
| 2040 | 22.72 | 20.20 | 20.54 | 16.44 | 190 |
| 2045 | 23.59 | 27.68 | 24.44 | 19.67 | 200 |

The results are evaluated based on the evolution of the technology mix and energy consumption. For this purpose, the building-specific individual results are aggregated and presented at the overall municipal level.

1) **Development of refurbishment measures**: Figure 5 shows the frequency distribution of renovation measures in the time range between 2023 and 2045. Four options (wall, window, roof, floor slab) are distinguished, and combinations of these are also possible if they are hygienically appropriate for the building (cf. [15]). By the year 2045, a total of 2216 renovation measures will be carried out on 962 buildings, which corresponds to an annual renovation rate of 1.39% and validates the model limitation of 2%. The results of the analysis show that renovation measures are carried out especially due to energy saving and reduction of $CO_2$ emissions. The energy renovation of buildings can help to reduce energy consumption by 8% and reduce environmental impact. There is an almost homogeneous distribution of measures over the years, with complete refurbishments tending to be carried out later; a significant factor in the decision to carry out a refurbishment measure is profitability and emissions reduction.

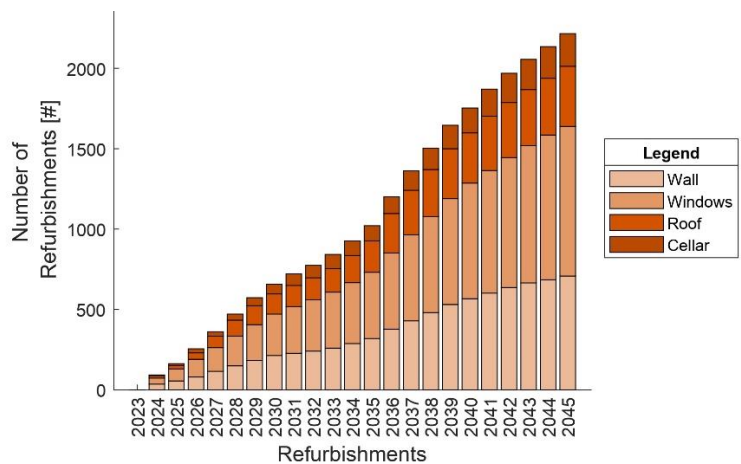

Figure 5. Frequency distribution of refurbishment

2) **Frequency distribution of heating systems**: The technological changes within the heating types, is shown in figure 6. In 2023, the heating stock is gas-dominated and has a share of 84%. However, during the transformation process, there is a shift towards more environmentally friendly technologies, which leads to a degressive trend of gas heating technology. This is driven by both rising operating costs and emission reductions from other technology options. Heat pumps in particular emerge as a key technology, accounting for 50% of all heating technologies in 2045. The distribution of air source heat pumps to ground source heat pumps is similar to the results from the literature (cf.[2],[16]). In this scenario, it was assumed that the heating network has already reached its maximum expansion in the SQ and therefore no new customers connect. Pellet heating and hybrid heating technologies have a crucial role in this scenario as a transitional technology, which scope between 2030 and 2040 will lead to a switch to more renewable energies in existing buildings with high energy demand or limited resource for renovations, as can be seen from the decline of the technology in 2045. Derived from Figure 6, the annual conversion rate of heating technologies is 4.5%, which corresponds to the maximum rate defined in the parameters.

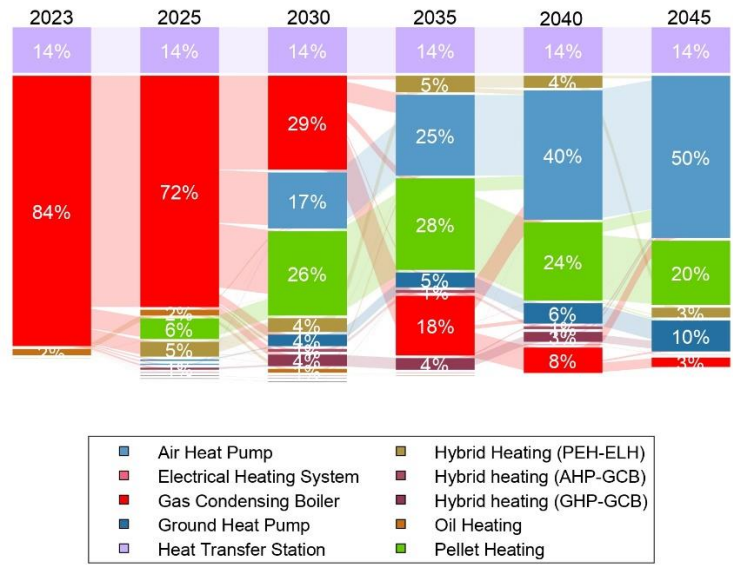

Figure 6. Frequency distribution of heating systems

3) **Installed power of the building systems**: Figure 7 shows the installed power of the different building technologies for the period under consideration, broken down by thermal (left column) and electrical systems (right column). In 2023, the installed heating load was 315 MW and a degressive trend in the maximum heating load can be seen over the years, which is due to the energy renovation of buildings. The trend shows a decrease of the installed power by 13,7% compared to the initial year. The energy refurbishment of buildings is an important aspect in the reduction of the heating load, as it reduces the heat demand of the buildings. Solar thermal systems, as described in DIN V 4701-10 ([17]), are not used to reduce the heating load, but as a support technology to improve the operating result. As already shown in Figure 6, the installed capacity of the heating technologies is mainly determined by gas condensing boilers. During the transformation process, there is an almost complete shift away from gas demand by 2045. For oil heating systems, this phase-out is already achievable in year 2035. Pellet heating systems have a maximum share of 28%, but account for 46% of the installed system capacity. This can be explained by the fact that this technology is especially advantageous for larger applications, due to the high investment costs, with low operating costs. Photovoltaic systems represented only 4 MW of capacity in 2023. In the following 22 years, there will be an a x18 increase increase due to local self-use and feed-in tariffs. The expansion of battery storage is no economicly suitable solution with the assumed prices and remuneration mechanisms over the entire period.

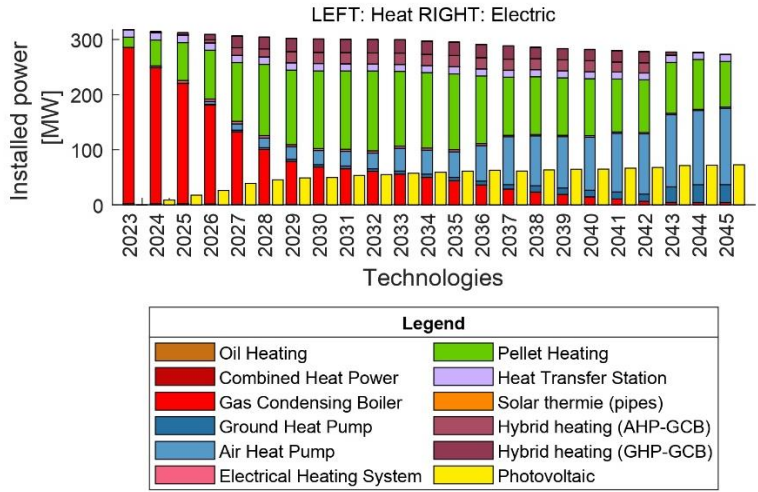

Figure 7. Installed power of the equipment technologies

4) **Energy balance** of the municipal energy system: In this study, the energy balance of the municipal energy system is examined and presented in Figure 8. The balance is divided into electrical and thermal energy to ensure a detailed insight into the energetic conditions. In the electrical balance, there is a significant increase in demand by 105%. The reason for this development is the strong addition of heat pumps. The increasing demand is partly compensated by decentralized generation from photovoltaic systems, although only a small proportion (approx. 24%) can be used by the residents themselves. This effect is related to electricity demand in the evening hours and electricity generation from photovoltaics in the midday hours. Figure 8 also provides an insight into the potentials of photovoltaic electricity generation, which will be further expanded in the later years of the time range investigated. However, effective utilization of these potentials depends on the application of storage technologies to enable photovoltaic power generation during midday hours to be used during evening hours. An important finding of the study is that heating demand can be reduced by up to 7.8% through refurbishment measures. This effect is enhanced by converting existing heating technologies to more efficient heat pumps. The results of the study thus illustrate that renovation measures can make an important contribution to reducing energy demand and thus help to achieve climate targets.

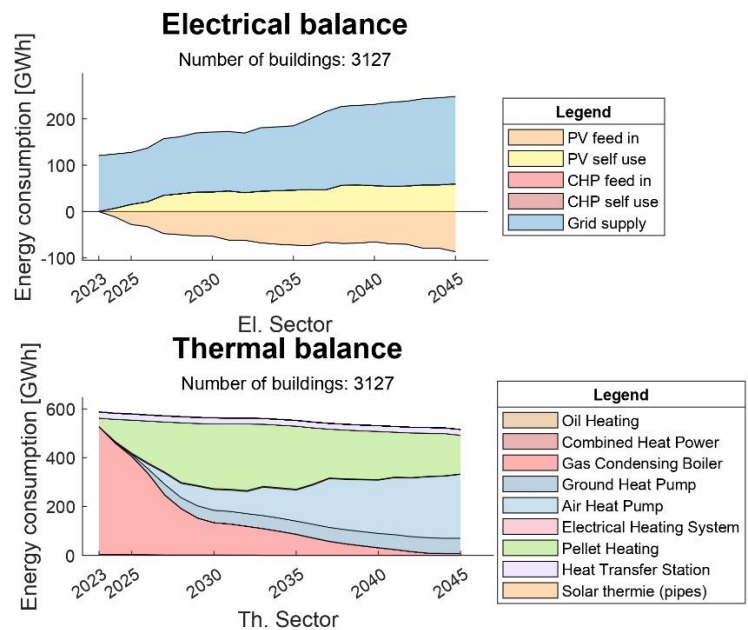

Figure 8. Energy Balance of the buildings

5) **Georeferenced development**: Figure 9 shows the georeferenced development of the heating distribution for the year 2045, the colors corresponding to the classification from Figure 8 Thermal Balance. It can be seen that buildings with sufficient open space use geothermal heat pumps (dark blue), due to the large area required for geothermal probes. Occasionally, initial gas customers can still be seen. In this analysis, the network operating costs for the continued operation of the infrastructure for fewer network customers were neglected. In addition, no clear end of the gas network was defined, which means that existing customers can continue to operate their gas heating systems. Pellet heating systems (green) can be seen in part in larger consumers, but also in buildings with a low renovation potential and a high energy density.

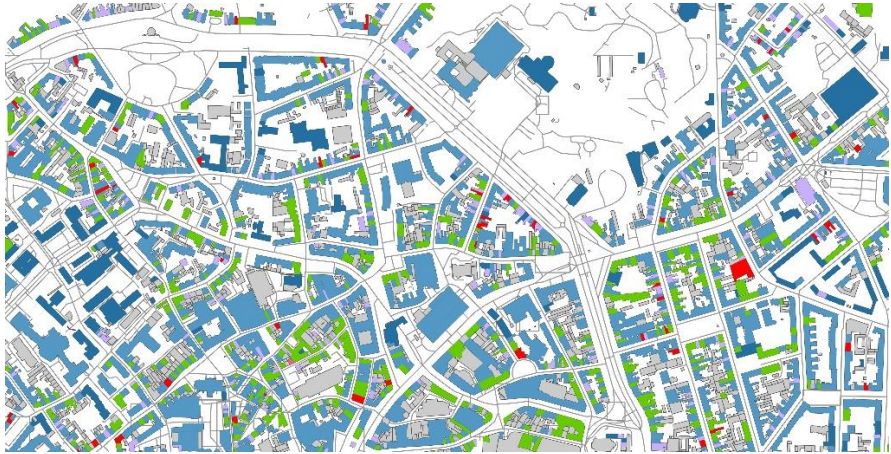

Figure 9. Heating technologies in 2045, legend according to Figure 8

6) **Development of costs**: The costs of the transformation process are shown in figure 10. For each technology, the investment, operation, maintenance and dismantling costs are aggregated. The figure is divided into heat generation technologies (left), electricity generation technologies (middle) and refurbishment costs (right). It can be seen that in the area of heat generation a significant reduction in costs of 47% has been achieved. This is due to two factors: First, there was a switch from gas heating to heat pumps. Second, the degressive development of energy prices, especially electricity prices, combined with a higher share of PV generation and own consumption, led to a reduction in costs. Total costs can be reduced by 43% compared to 2023 as a result of these measures. It can be seen that the cost of electricity increases mainly in the years up to 2030 due to the expansion of photovoltaic installations. In later years, these plants can even be seen to generate positive returns (2040). The cost of refurbishment is barely visible in this graph, which is due to the fact that refurbishment is only a small part of the very high operating costs of heating technologies. In the optimisation, the benefits of the refurbishment are compared to the costs, which means that not every refurbishment is economical for every building.

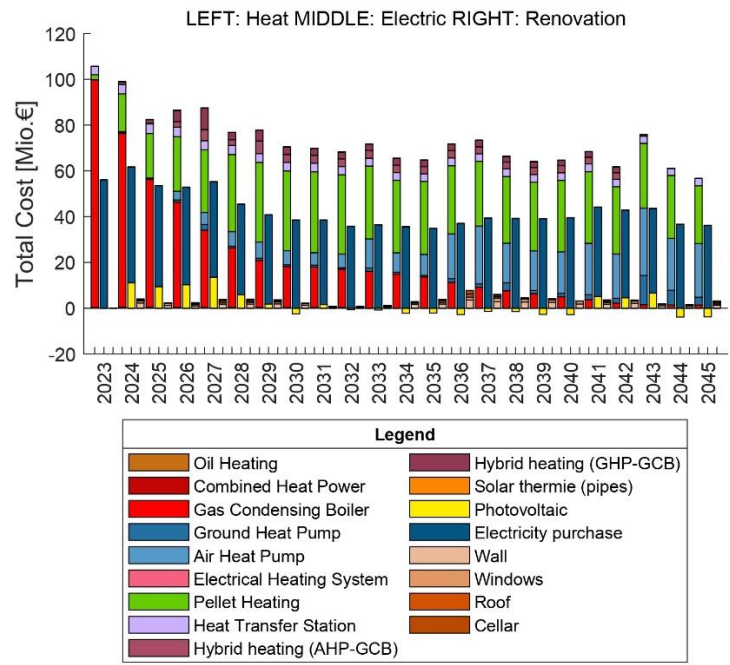

Figure 10. Investment costs and operating costs of end customers

7) **Development of emissions**: Figure 11 shows the emission quantities for the three areas heat, electricity and refurbishment, divided in the same way as figure 10. The emissions include scope 1-3 ([18]). In 2023, gas combustion is the dominant factor with a share of 72%. However, this emission factor can be reduced by a gradual change in supply and the addition of synthetic methane, which is why these emissions are < 1% in 2045. Similar transformation processes can also be observed in the area of electricity supply. Overall, an emission reduction of 97% can be achieved. The remaining emissions result from the emissions during the production and recycling of the technology, such as the energy sources pellets and electricity. Emissions of 37.5g $CO_2$eq./kWh were assumed for electricity mix in 2045 ([9]).

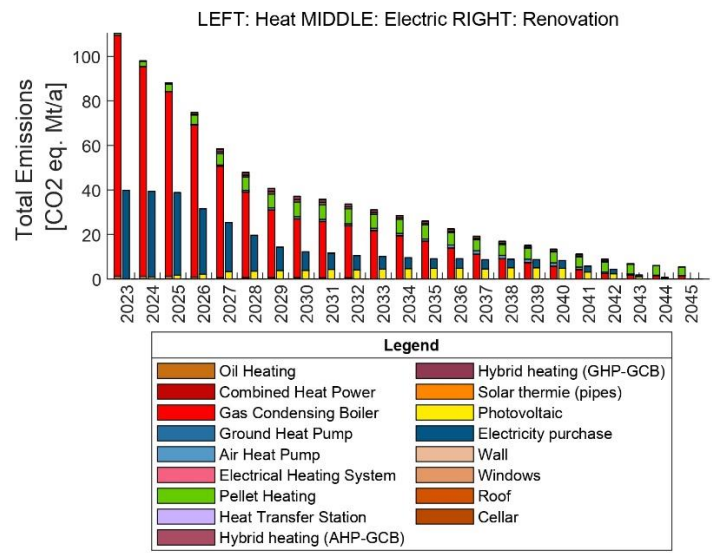

Figure 11. Emissions for the installation, operation and dismantling of end-user plants

## 4. CONCLUSION

Reducing carbon emissions in the energy sector necessitates a transition path that describes the conversion of the existing system to alternative energy sources. In the process, municipalities are an important part in the transition due to their high energy demand and proximity to citizens. Especially in the heating sector, the share of renewable energy is still low. This paper presents a model that analyzes cost-based transitions in the municipal sector. This model accounts for temporally dynamic factors to demonstrate their impact on the multi energy system's design. A partial assessment of the dynamics could be verified by applying the model. This resulted in revealing a possible transformation path for the municipal energy system. On the whole, it can be concluded that introducing renewable energy and expanding energy efficiency measures can help the municipal region under consideration make significant progress in decarbonizing the building sector. By presenting figures and trends, it is evident that the use of innovative technologies can facilitate a successful transformation of the building sector. The results of this study can form the basis for further research and decision-making in the field of building retrofits. It's crucial to formulate retrofit strategies for the future that are environmentally and economically sustainable and that contribute to achieve climate goals. The analysis of the municipal energy system indicates a substantial increase in heat pumps in the coming years, leading to a significant demand for electrical energy. Conversely, decentralized generation from photovoltaic systems is also on the rise, albeit only a small proportion (approx. 24%) of which is usable by the residents themselves. These effects are associated with the electricity demand in the evening and photovoltaic electricity generation at noon. Renovations can reduce heating demand by up to 7.8%, and the conversion of existing heating technologies to heat pumps can also play a crucial role in the decarbonization of the heating sector. Future studies should include price uncertainties to derive robust system design decisions. In addition, a purely cost-driven optimization approach should be questioned and extended to include social aspects.